\documentstyle[11pt]{article}
\begin{document}

\title{The QCD corrections to scalar top pair production
in photon-photon collisions
\footnote{The project supported by National
Natural Science Foundation of China}}

\author{{
  CHANG Chao-Hsi$^{a,b}$, HAN Liang$^{c}$, MA Wen-Gan$^{a,b,c}$ and
  YU Zeng-Hui$^{c,d}$}\\
{\small $^{a}$CCAST (World Laboratory), P.O.Box 8730, Beijing 100080,
China.}\footnote{Not mailing address.}\\
{\small $^{b}$Institute of Theoretical Physics, Academia Sinica,}\\
{\small P.O.Box 2735, Beijing 100080, China.}\\
{\small $^{c}$Department of Modern Physics, University of Science
and Technology}\\
{\small of China (USTC), Hefei, Anhui 230027, China.}\\
{\small $^{d}$Institut f$\ddot{u}$r Theoretische Physik, Universit$\ddot{a}$t
        Wien, A-1090 Vienna, Austria} }
\date{}
\maketitle

\begin{center}\begin{minipage}{100mm}

\vskip 5mm
\begin{center} {\bf Abstract}\end{center}
\baselineskip 0.3in
{The complete QCD corrections to the scalar top quark pair production via
$\gamma \gamma$ fusion in parent $e^{+}e^{-}$ collider are calculated in
the Supersymmetric Model. The various origins of corrections are discussed.
We find that the correction due to gluino exchange is smaller than that due to
conventional gluonic contribution, and the contribution due to
squark quartic interactions is much smaller. Moreover, the
dependences of the corrections on the masses
and the mixing angle of the supersymmetric particles
are also investigated. } \\

\vskip 5mm
{PACS number(s): 14.80.Ly, 12.15.Lk, 12.60.Jv}
\end{minipage}
\end{center}

\newpage
\noindent
{\Large{\bf 1.Introduction}}
\vskip 5mm

\begin{large}
\baselineskip 0.35in

  The Standard Model (SM) \cite{glash} \cite{higgs} has achieved
great successes in describing all the available experiment
data pertaining to the strong, weak and electromagnetic interaction
phenomena up to a few percentage level quantitatively. Only the elementary
Higgs boson, required strictly by the Standard Model for spontaneous
symmetry breaking, remains to be found. At present the Supersymmetric
extended Model (MSSM)\cite{haber} is regarded as one of the most attractive
model. Apart from containing a large number of experimental consequences,
the supersymmetric theory is able to solve various theoretical problems,
such as the fact that SUSY maybe provides an elegant way to construct
the huge hierarchy between the electroweak symmetry-breaking scale and
the grand unification scales.

One of the most important consequence of the SUSY model is to predict
the existence of scalar partners to all existing fermions in the SM.
In the model two coloured scalar quarks (squarks)
$\tilde q_{L}$ and $\tilde q_{R}$ are required as partners of the
standard Dirac quarks. To probe the existence of these squarks, to look
for pair production $\tilde{q} \bar{\tilde{q}}$ would be a very promising.
The pair production may have a comparatively
large cross section, depending on the masses
of the squark. Furthermore, if both pair-produced squarks decay
via $\tilde q \rightarrow q \tilde{\gamma}$, the signature
with hard acoplanar quark jets, may be quite
good for detection, allowing identification
of individual events which would not occur in the Standard Model.

  Since $\tilde q_{L}$ and $\tilde q_{R}$ will turn to their
mass eigenstates by mixing with each other if they are produced,
and the mixing size is proportional to the mass of the
related ordinary quark \cite{sa}.
As the top quark is very heavy, the scalar partners of top quark (stop quark)
are expected to be strongly mixed. Therefore, one of
the two mass eigenstates of stop quarks
$\tilde t_{1}$ and $\tilde t_{2}$, which are mixtures of
the left and right weak eigenstates of the stops
$\tilde q_{L}$ and $\tilde q_{R}$, may be comparative light
(the lightest one of the all squarks) i.e. we may suppose the one of
the stop mass eigenstates $\tilde t_{1}$ is much lighter than
the other one $\tilde t_{2}$. As a result, $\tilde t_{1}$ is very possible
to be discovered in a relatively low energy range.

  The Next-generation Linear Collider (NLC) is designed to look for the
evidences of Higgs and other new particles beyond the Standard Model. To detect
the existence of the heavy squarks, $e^{+}e^{-}$ collision has advantage
over $pp$ or $p \bar p$ collision with its cleaner background. Squark pair
production produced by $e^{+}e^{-}$ annihilation has been studied thoroughly,
both at tree level \cite{sb} and at next-to-leading \cite{sc}. The overall
consideration of producing a scalar top pair via $\gamma \gamma$ collision is
still absent and its production rate may be greater than that by the $e^+e^-$
annihilation as the later has an `s-channel suppression' due to the virtual
photon propagator when the scalar top quark is so heavy.
Moreover, in the process of squark pair production via photon
fusion, the final particles should not be two different squark mass
eigenstates, because it is merely electromagnetic interaction phenomenon
and no weak interaction is involved at the tree-level,
other than in the production of $e^{+}e^{-} \rightarrow
\tilde{q}_{i} \bar{\tilde{q}}_{j}$ $(i,j=1,2)$
there must be some contribution from Z boson exchange.
Therefore, $\gamma \gamma$
collision provides another interesting mode of producing squark pair
which is worthwhile to investigate. Furthermore, with the advent of new
collider technique, the NLC is not only able to work as $e^{+}e^{-}$ collider,
but also to turn the high energy electron-positron beams into the Compton
backscattering energetic photon beams with high efficiency \cite{sd}.
The luminosity and energy of the photon collider can be expected to be
comparable to those of the primary $e^{+}e^{-}$ collision. Therefore, with
the solid support of new experimental techniques, there is possibility to
yield a scaler top quark pair production directly via the high energy
photon collision.

  From the above argument we can see that the process of squark pair
production via photon-photon collisions would be interesting
and will be accessible experimentally. We know that the study for this
process only at the tree-level is far from enough in high precision. In
this work, we will investigate the 1-loop QCD corrections up to the order
$O(\alpha_s)$, including the conventional ones (virtual and real gluon
exchange) and those of supersymmetric partners. The paper is organized
as follows: In Sec.II the analytical formula of tree level is presented.
The different origins of supersymmetry QCD corrections are discussed in
Sec.III. In Sec.IV, the numerical results are depicted, and some discussions
are given. Finally, the conclusions are made. In the Appendix, the form
factors appearing in the cross section of Sec.III are
listed.\\

\vskip 5mm
\noindent
\begin{flushleft} {\bf II. Tree level result} \end{flushleft}
\vskip 5mm
  In this work, we denote the reaction of stop quark pair production
via photon-photon collision as
$$
\gamma (p_3, \mu) \gamma (p_4, \nu) \longrightarrow \tilde{t}_i (p_1)
                                                    \bar{\tilde{t}}_i (p_2)
\eqno{(2.1)}
$$
where $p_1$ and $p_2$ represent the momenta of the outgoing scalar top-quark and
its anti-particle, $p_3$ and $p_4$ describe  the momenta of the two incoming
photons.

The Feynman diagrams for the process at the tree level are shown in Fig.1-a,
and the relevant Feynman rules can be found in \cite{sa}. The corresponding
Lorentz invariant matrix element of the lowest order for the reaction
$\gamma \gamma \rightarrow \tilde{t}_i \bar{\tilde{t}}_i$ is written as
$$
M_0 = M_0^{\hat{t}} + M_0^{\hat{u}} + M_0^{\hat{q}}
\eqno{(2.2)}
$$
where $M_0^{\hat{t}}$, $M_0^{\hat{u}}$ and $M_0^{\hat{q}}$ represent the
amplitudes of the t-channel, u-channel and quartic coupling diagrams
respectively. The explicit expressions can be given as
$$
M_0^{\hat{t}} = \frac{4 i e^2 Q_{\tilde{t}}^2}{\hat{t} - m_{\tilde{t}_i}^2}
                \epsilon_{\mu}(p_3) \epsilon_{\nu}(p_4)~p_1^{\mu}~p_2^{\nu}
$$
$$
M_0^{\hat{u}} = \frac{4 i e^2 Q_{\tilde{t}}^2}{\hat{u} - m_{\tilde{t}_i}^2}
                \epsilon_{\mu}(p_3) \epsilon_{\nu}(p_4)~p_1^{\nu}~p_2^{\mu}
$$
$$
M_0^{\hat{q}} = 2 i e^2 Q_{\tilde{t}}^2  g^{\mu\nu} \epsilon_{\mu}(p_3) \epsilon_{\nu}(p_4)
\eqno{(2.3)}
$$
 The Mandelstam variables $\hat{t}$ and $\hat{u}$ are defined as $\hat{t}=(p_1
- p_3)^2,~\hat{u}=(p_1 - p_4)^2$. $m_{\tilde{t}_i}~(i=1,2)$ and $Q_{\tilde{t}}$
denote the masses and the charge of scalar top quarks respectively.

  The cross section at tree level can be depicted as
$$
\hat{\sigma}_{B}[\gamma \gamma \rightarrow \tilde{t}_i \bar{\tilde{t}}_i] =
   \frac{4 \pi \alpha^2}{\hat{s}}Q_{\tilde{t}}^4~N_C~\beta~
   \{ 1 + \frac{8 \mu^4}{1 - \beta^2} +
      \frac{2 \mu^2 (1 - 2 \mu^2)}{\beta} \log{w} \}
\eqno{(2.4)}
$$
Here $\mu^2=m_{\tilde{t}_i}^2/\hat{s}$, the velocity of the scalar top quarks
is defined as $\beta=\sqrt{1 - 4 m_{\tilde{t}_i}^2/\hat{s}}$. The kinematical
variable $w$ is defined as $w = (1 - \beta)/(1 + \beta)$. $N_C = 3$ is
the number of colors.

\vskip 5mm
\noindent
\begin{flushleft} {\bf III. SUSY QCD corrections} \end{flushleft}

Since the physical squarks are the mixture of the left- and right-handed
weak eigenstates, the stop quark mass
eigenstates $\tilde{t}_1$ and $\tilde{t}_2$
can be expressed by
$$
\tilde{t}_1 = \cos{\theta_{\tilde{t}}}~\tilde{t}_L -
              \sin{\theta_{\tilde{t}}}~\tilde{t}_R, ~~~~
\tilde{t}_2 = \sin{\theta_{\tilde{t}}}~\tilde{t}_L +
              \cos{\theta_{\tilde{t}}}~\tilde{t}_R
\eqno{(3.1)}
$$
where $\theta_{\tilde{t}}$ is the scalar top mixing angle which transforms
the scalar top mass eigenstates $\tilde{t}_i$ (i=1,2) to the weak eigenstates
$\tilde{t}_L$ and $\tilde{t}_R$. We assume that the scalar top quark mass
eigenstates split heavily, namely $m_{\tilde{t}_1} << m_{\tilde{t}_2}$.
Therefore, in this work we constrain our attention only on the calculation
of the $\tilde{t}_1 \bar{\tilde{t}}_1$ pair production.

  The Feynman diagrams for the process $\gamma \gamma \rightarrow \tilde{t}_1
\bar{\tilde{t}}_1$ including all the QCD corrections up to the order
$O(\alpha_s)$ are depicted in Fig.1. The total cross section with the
supersymmetric QCD corrections corresponding to Fig.1 can be denoted as

$$
\hat{\sigma} = \hat{\sigma}_B + \hat{\sigma}_{soft} +
                 \delta \hat{\sigma}^{(1-loop)}
\eqno{(3.2)}
$$
where $\hat{\sigma}_{soft}$ and $\delta \hat{\sigma}^{(1-loop)}$ represent
the cross section of soft gluon emission and virtual one-loop QCD corrections
respectively.

  The one-loop supersymmetric QCD corrections originate from three different
sources: the conventional virtual gluonic corrections (see Fig.1-b), those
due to the gluino exchange and those due to squark quartic interactions
(see Fig.1-d). In order to show the virtual effects from the different sources,
we divide the loop corrections into three groups correspondingly. In the
calculation, the dimensional regularization and on-mass-shell
scheme (OMS) \cite{se} are adopted in doing renormalization, whereas when
considering the effects from the soft gluons
infrared divergences being
regularized by introducing a small no-zero gluon mass is adopted.

\vskip 5mm
\begin{flushleft} {\bf 3.1 Related self-energy and counterterms.}
\end{flushleft}

The unrenormalized self-energy of the scalar top $\tilde{t}_1$ can be
written as the summation of three parts.
$$
\Sigma (p^2) = \Sigma^{(g)}_{vir}(p^2) + \Sigma^{(\tilde{g})}(p^2) +
               \Sigma^{(\tilde{t})}(p^2)
\eqno{(3.3)}
$$
where $\Sigma^{(g)}_{vir}$, $\Sigma^{(\tilde{g})}$ and $\Sigma^{(\tilde{t})}$
denote the scalar top self-energy parts contributed by the diagrams
the virtual gluon(Fig.1-b1),
gluino exchange(Fig.1-d1) and the scalar top quartic
interactions(Fig.1-d5)
respectively. The squark quartic interactions are involved by introducing
superpotential of the scalar matter fields \cite{sa}. Imposing the OMS
renormalization conditions \cite{se}, one can get the related renormalization
constants to $O(\alpha_s)$ order with the formulas
$$
\delta M^{2} =\delta M^{2(g)}+ \delta M^{2(\tilde{g})}+ \delta M^{2(\tilde{t})}=
            \tilde{Re} \Sigma (m_{\tilde{t}_1}^2),
$$
$$
\delta Z = \delta Z^{(g)} + \delta Z^{(\tilde{g})} + \delta Z^{(\tilde{t})}=
           -\tilde{Re} \frac{\partial \Sigma (p^2)}
                                        {\partial p^2}
                                        |_{p^2=m_{\tilde{t}_1}^2},
\eqno{(3.4)}
$$
where $\tilde{Re}$ only takes the real part of the loop integral functions
appearing in the self-energies. All the definitions of loop integral functions
$A$, $B$, $C$ and $D$ used in the paper can be found in Ref.\cite{sf}.

    The three contribution parts of the scalar top $\tilde{t}_1$ self-energy
can be written explicitly as
$$
i \Sigma^{(g)}_{vir}(p^2) = -i \frac{g_s^2 C_F}{16 \pi^2} \left \{
                      A_0[m_{\tilde{t}_1}] +
                      4 p^2 (B_0 + B_1)[p^2, m_g, m_{\tilde{t}_1}] +
                      m_g^2 B_0[p^2, m_g, m_{\tilde{t}_1}] \right \},
\eqno{(3.5)}
$$
$$
i \Sigma^{(\tilde{g})}(p^2) = i \frac{g_s^2 C_F}{16 \pi^2} d
  \{ A_0[m_t] -
     (m_{\tilde{g}}^2 + m_t m_{\tilde{g}} \sin{2 \theta_{\tilde{t}})}
     B_0[p^2, m_{\tilde{g}}, m_t] - p^2 B_1[p^2, m_{\tilde{g}}, m_t]
  \},
\eqno{(3.6)}
$$
$$
i \Sigma^{(\tilde{t})}(p^2) = -i \frac{g_s^2}{12 \pi^2}
        \{ A_0[m_{\tilde{t}_1}^2]~cos^2 2 \theta_{\tilde{t}} +
           A_0[m_{\tilde{t}_2}^2]~sin^2 2 \theta_{\tilde{t}} \},
\eqno{(3.7)}
$$
where $m_g$ is denoted as the gluon mass and $d=4-\epsilon$ is the space-time
dimension.

    Note that the self-energy part from stop quartic interaction
diagram(Fig.1-d5) expressed in Eq.(3.7), is independent on its momentum
squared $p^2$. This means the corresponding part of the squark wave function
renormalization constant $\delta Z^{(\tilde{t})} = 0$. The other related
renormalization constants $\delta Z^{(g)}$, $\delta Z^{(\tilde{g})}$,
$\delta M^{2(g)}$,$\delta M^{2(\tilde{g})}$ and $\delta M^{2(\tilde{t})}$
can be obtained by using the expressions of $Eq.(3.3 \sim 3.7)$.

\begin{flushleft} {\bf 3.2 Renormalized one-loop corrections.} \end{flushleft}

    The renormalized one-loop matrix element involves the contributions from
all the SUSY QCD one-loop self-energy, vertex, box and quartic interaction
diagrams(shown in $Fig.1-(b1 \sim b12)$ and $Fig.1-(d1 \sim d7)$) and their
relevant counterterms. It can be written in the form as

$$
\begin{array} {lll}
\delta M^{(1-loop)} &=& \delta M^{(g)}_{vir} + \delta M^{(\tilde{g})} +
                        \delta M^{(\tilde{t})} \\ [3mm]
                    &=& i e^2 Q_{\tilde{t}}^2 \epsilon_{\mu}(p_3)
                      \epsilon_{\nu}(p_4)
                        [ f_1 ~ g^{\mu \nu} +
                          f_2 ~ p_1^{\mu} p_1^{\nu} +
                          f_3 ~ p_1^{\mu} p_2^{\nu} +
                          f_4 ~ p_2^{\mu} p_1^{\nu} +
                          f_5 ~ p_2^{\mu} p_2^{\nu} ],
\end{array}
\eqno{(3.8)}
$$
with
$$
f_i = f_{vir, i}^{(g)} + f_i^{(\tilde{g})} + f_i^{(\tilde{t})},
\eqno{(3.9)}
$$
where $\delta M^{(g)}_{vir}$ represents the gluon correction part of the
one-loop renormalized amplitude, and $\delta M^{(\tilde{g})}$,
$\delta M^{(\tilde{t})}$ represent the gluino exchange and the squark
quartic interactions contribution respectively. All the renormalized
self-energy, vertex, box and quartic interaction loop corrections due
to the different sources, are summarized in Lorentz invariant
factors $f_{vir, i}^{(g)}$, $f_i^{(\tilde{g})}$ and $f_i^{(\tilde{t})}~~
(i=1,5)$, respectively. By using this formulation we are to show a
clear distinction among those three different origins of supersymmetry
QCD corrections. The explicit expressions of form factors can be found
in Appendix.

    Now we can achieve the one-loop SUSY QCD corrections of the cross section
for this subprocess in unpolarized photon collisions.

\begin{eqnarray*}
 \delta \hat{\sigma}^{(1-loop)}(\hat{s}) &=& \frac{N_C}{16 \pi \hat{s}^2}
             \int_{\hat{t}^{-}}^{\hat{t}^{+}} d\hat{t}~
             2 Re {\bar{\sum\limits_{spins}^{}}} \left( M_{0}^{\dag} \cdot
             \delta M^{(1-loop)} \right) \\
         &=& \delta \hat{\sigma}^{(g)}_{vir} +
             \delta \hat{\sigma}^{(\tilde{g})} +
             \delta \hat{\sigma}^{(\tilde{t})} \\
         &=& \delta \hat{\sigma}^{(g)}_{vir} +
             \delta \hat{\sigma}^{(\tilde{g} + \tilde{t})},
\end{eqnarray*}
$$
\eqno{(3.10)}
$$
where $\hat{t}^\pm=(m_{\tilde{t}_1}^2-\frac{1}{2}\hat{s})\pm\frac{1}{2}\hat{s}
\beta$. The bar over the sum recalls averaging over initial spins.
The correction $\delta \hat{\sigma}^{(\tilde{g} + \tilde{t})} =
\delta \hat{\sigma}^{(\tilde{g})} + \delta \hat{\sigma}^{(\tilde{t})}$ will be
used to denote the corrections arising from the pure supersymmetric particle
contribution.

\vskip 5mm
\begin{flushleft} {\bf 3.3 Real gluon emission} \end{flushleft}

   The real gluon emission diagrams for the process $\gamma \gamma \rightarrow
\tilde{t}_1 \bar{\tilde{t}}_1 g$ are drawn in $Fig.1-c1 \sim c7$. The
cross section of real gluon emission is deduced as
$$
\sigma_{real} = \frac{N_c}{4} \frac{1}{2 \hat{s} (2 \pi)^5}
                  \int_{}^{}
                  \frac{d^3 \vec{p}_1}{2 E_1}
                  \frac{d^3 \vec{p}_2}{2 E_2}
                  \frac{d^3 \vec{k}}{2 \omega}
                  \delta^4 (p_3 + p_4 - p_1 - p_2 - k)
                  {\bar{\sum\limits_{spins}^{}}} |M_{real}|^2
\eqno{(3.11a)}
$$
with
\begin{eqnarray*}
{\bar{\sum\limits_{spins}^{}}} |M_{real}|^2 &=& C_F g_s^2
           {\bar{\sum\limits_{spins}^{}}} \left(
           f_{real}^{ \hat{t} \hat{t}} \cdot |M_0^{\hat{t}}|^2 +
           f_{real}^{ \hat{u} \hat{u}} \cdot |M_0^{\hat{u}}|^2 +
           f_{real}^{ \hat{q} \hat{q}} \cdot |M_0^{\hat{q}}|^2 \right. \\
&+& 2 f_{real}^{ \hat{t} \hat{u}} \cdot Re ( M_0^{\hat{t} \dag}
                                                    M_0^{\hat{u}} )
                + 2 f_{real}^{ \hat{q} \hat{t}} \cdot Re ( M_0^{\hat{q} \dag}
                                                    M_0^{\hat{t}} )
                + \left. 2 f_{real}^{ \hat{q}\hat{u}} \cdot  Re
( M_0^{\hat{q} \dag} M_0^{\hat{u}}) \right)
\end{eqnarray*}
$$
\eqno{(3.11b)}
$$
where $\vec{k}$ and $\omega$ denote the momentum and the energy of the outgoing
gluon respectively, and $M_0^{a}~(a=\hat{t},\hat{u},\hat{q})$ are the
invariant matrix elements of tree level defined above. The factors
$f_{real}^{a b}~(a,b=\hat{t}, \hat{u}, \hat{q})$ can be found in Appendix.

  In the region around $k \sim 0$, the cross section in Eq.(3.11),
corresponding to the cross
section for a soft gluon being emitted, may be approximately alternated as
$$
\left ( \frac{d \sigma}{d \Omega} \right )_{soft} =
-\left ( \frac{d \sigma}{d \Omega} \right )_{B} \frac{g_s^2 C_F}{(2 \pi)^3}
[I_{ii} + I_{ij}]
\eqno{(3.12)}
$$
where
$$
I_{ii} = 2 \pi \left \{ 2 log[\frac{2 \Delta E}{m_g}] + \frac{log[w]}{\beta} \right \}
\eqno{(3.13a)}
$$
$$
I_{ij} = 2 \pi \frac{1 - 2 \mu^2}{\beta} \left \{
                     2 log[w] log[\frac{2 \Delta E}{m_g}] +
                     2 Li_2[\frac{2 \beta}{1 + \beta}] +
                     \frac{1}{2} log^2[w] \right \}
\eqno{(3.13b)}
$$
Here a small soft gluon energy cut $\Delta E$ is introduced.
$Li_2[x]$ is the Spence function. The definition of $\beta$ is the same
as in Eq.(2.4). The expressions Eq.(3.12) and Eq.(3.13a,b) are similar to
those as described in Ref.\cite{se} for a soft photon being emitted,
that just is a check of Eq.(3.11) for the real gluon emission
in a certain sense.
However, for the cancellation of the infrared divergences
due to gluons, it is sufficient that only the
soft gluons , i.e. gluons with energy $\omega \leq \Delta E$, are relevant,
where $\Delta E$ is small, which may be set to be smaller than
all relevant energy scales in concerned problem. Therefore, we adopt
the soft gluon approximation method with a cut $\Delta E$
in numerical calculations for the soft gluon emission corrections.
With this approach the radiation from internal colour lines or
quartic vertices does not lead to IR-singularities and can be neglected in
this approximation.

   The correction to the total cross section from the gluonic QCD contributions
should be expressed as below

$$
\delta \hat{\sigma}^{(g)} = \delta \hat{\sigma}^{(g)}_{vir} +
                            \hat{\sigma}_{soft}
\eqno{(3.14)}
$$

\vskip 5mm
\noindent
\begin{flushleft} {\bf IV. Numerical results and discussions}
\end{flushleft}
\vskip 5mm

In the numerical evaluation, we take the value of the top-quark mass to be
$m_t=175~GeV$, $~\alpha = 1/128$ and use the one-loop running coupling
constant $\alpha_s$. We assume $m_{\tilde{t}_1} < m_{\tilde{t}_2}$ for the
masses of the stop quark mass eigenstates. The present constraint from
experiments\cite{sg} is that the lighter scalar top mass
eigenstate $m_{\tilde{t}_1}$ should be heavier than 50 GeV.

The corrections to the cross section of the process $\gamma \gamma
\rightarrow \tilde{t}_1 \bar{\tilde{t}}_1$ as a function of the center-of-mass
$\sqrt{\hat{s}}$ are studied with the values of the involved SUSY parameters
given below
$$
m_{\tilde{t}_1}=200~GeV,~~m_{\tilde{t}_2}=550~GeV,~~m_{\tilde{g}}=250~GeV,~
~cos \theta_{\tilde{t}} = 0.7
\eqno{(4.1)}
$$
The results are depicted in Fig.2(a). These results of numerical calculations
show
(i) Both corrections $\delta \hat{\sigma}^{(g)}$ and
$\delta \hat{\sigma}^{(\tilde{g}+\tilde{t})}$ reach their maximal values
around the threshold of $\tilde{t}_1 \bar{\tilde{t}}_1$ pair production
and vary from positive to negative value as increasing $\sqrt{\hat{s}}$.
The former varies severely and in general is several times higher than
the later.
(ii) The relative corrections due to the gluon contribution can even
reach $+50\%$ when c.m.s energy $\sqrt{\hat{s}}$ is just
above the pair threshold, whereas the maximal value of
$\delta \hat{\sigma}^{(\tilde{g}+\tilde{t})}/\hat{\sigma}_B$ is only
about $+10\%$. Therefore, one may see the fact that
the SUSY QCD corrections are dominated by the
gluon contributions.

The $\tilde{t}_1 \bar{\tilde{t}}_1$ pair production via photon-photon fusion
is only a subprocess of the parent $e^{+}e^{-}$ linear collider. It is easy
to obtain the total cross section for the scalar top quark pair production via
photon fusion in the $e^{+}e^{-}$ collider, by simply folding the cross
section of the subprocess $\hat{\sigma} (\gamma\gamma \rightarrow
\tilde{t}_1 \bar{\tilde{t}}_1)$ with the photon luminosity.
$$
\sigma(s)=\int_{2 m_{\tilde{t}_1}/\sqrt{s} }^{x_{max}}
dz \frac{dL_{\gamma\gamma}}{dz} \hat{\sigma}(\gamma\gamma\rightarrow
 \tilde{t}_1 \bar{\tilde{t}}_1\hskip 3mm at \hskip 3mm \hat{s}=z^2 s),
\eqno{(4.2)}
$$
where $\sqrt{s}$ and $\sqrt{\hat{s}}$ are the $e^{+}e^{-}$ and $\gamma \gamma$
center-of-mass energies respectively and $dL_{\gamma\gamma}/dz$
is the distribution of photon luminosity, which is expressed as
$$
\frac{dL_{\gamma\gamma}}{dz}=2z\int_{z^2/x_{max}}^{x_{max}}
 \frac{dx}{x} f_{\gamma/e}(x)f_{\gamma/e}(z^2/x),
\eqno{(4.3)}
$$
where $f_{\gamma/e}$ is the photon structure function of the electron beam
\cite{sd,sh}. We take the structure function of the photon produced by
the most promising Compton backscattering as \cite{sd,si}

\[
f^{Comp}_{\gamma/e} =\left\{
\begin{array}{cl}
\frac{1}{1.8397} \left(1-x+\frac{1}{1-x}-\frac{4x}{x_{i}(1-x)}+\frac{4 x^2}{x_{i}^2 (1-x)^2} \right),
& {\rm for}~x~<~0.83, x_{i}=2(1+\sqrt{2}) \\
0, & {\rm for}~x~>~0.83.
\end{array}
\right.
\]
$$
\eqno{(4.4)}
$$

The cross sections of the process $e^{+}e^{-} \rightarrow \gamma \gamma
\rightarrow \tilde{t}_1\bar{\tilde{t}}_1$ versus $\sqrt{s}$, with the same
condition of Eq.(4.1), are depicted in Fig.2(b). There the magnitude of the
corrections is also significant for this process and the sign of corrections
goes from positive to negative with the increasing of $\sqrt{s}$.
The total cross section involving SUSY QCD corrections shows the weak
dependence on the parent c.m.s energy when $\sqrt{s}>1~TeV$.
Furthermore, by analysing our numerical data,
the whole correction $\delta \sigma/\sigma_B$ can reach $+35\%$.

  In Fig.3 the concerned corrections are plotted as functions of the light
scalar top mass $m_{\tilde{t}_1}$, with $\sqrt{\hat{s}}=1000~GeV$,
$m_{\tilde{t}_2}=550~GeV$, $m_{\tilde{g}}=250~GeV$ and
$cos \theta_{\tilde{t}} = 0.7$. In most regions of the value
$m_{\tilde{t}_1}$, the corrections due to gluonic contributions are much
larger than those due to the scalar top quarks and gluino contributions.
Around the region of $m_{\tilde{t}_1} \sim 500~ GeV$ where $\sqrt{\hat{s}}$
just reaches its pair production threshold, the gluonic corrections become
positive, whereas they are negative when $\sqrt{\hat{s}}$ is far above the
threshold.

  In Fig.4 the concerned corrections with various $m_{\tilde{t}_2}$
and $\sqrt{\hat{s}}=500~GeV$, $m_{\tilde{t}_1}=200~GeV$,
$m_{\tilde{g}}=250~GeV$, $cos\theta_{\tilde{t}}=0.7$ as well, are depicted.
Since the heavy stop mass $m_{\tilde{t}_2}$
only appears in the amplitudes of the diagrams involving the squark quartic
vertex, the dependence of the corrections on $m_{\tilde{t}_2}$
manifests only in the correction $\delta \hat{\sigma}^{(\tilde{t})}$. The
correction $\delta \hat{\sigma}^{(\tilde{t})}/\hat{\sigma}_B$
approaches to zero with the increment of $m_{\tilde{t}_2}$ and the maximal
correction is only about $-1\%$. Note that when $m_{\tilde{t}_2} = 550 ~GeV$,
the relative correction $\delta \hat{\sigma}^{(\tilde{t})}/\hat{\sigma}_B$
is only $-0.2\%$, whereas in Fig.2(a) the correction
$\delta \hat{\sigma}^{(\tilde{t} + \tilde{g})}/\hat{\sigma}_B$
with the same condition is about $+4\%$.
That shows $\delta \hat{\sigma}^{(\tilde{t})}$ is always smaller than both
$\delta \hat{\sigma}^{(g)}$ and $\delta \hat{\sigma}^{(\tilde{g})}$.
Moreover, the curve of the correction due to the squark quartic
interactions shown in Fig.4, becomes rather flat in the large $m_{\tilde{t}_2}$
region, which implies the dependence of $\delta \hat{\sigma}^{(\tilde{t})}$
on heavy $\tilde{t}_2$ is very weak.

  The gluino exchange contribution to the cross section is calculated as a
function of $m_{\tilde{g}}$. The results can be seen in Fig.5, where the
parameters are set as $\sqrt{\hat{s}}=500~GeV$, $m_{\tilde{t}_1}=200~GeV$,
$m_{\tilde{t}_2}=550~GeV$ and $cos \theta_{\tilde{t}}=0.7$.
The curve of the correction
$\delta \hat{\sigma}^{(\tilde{g})}/\hat{\sigma}_B$ is nearly parallel to
that of the gluon and stop contribution
$\delta \hat{\sigma}^{(g+\tilde{t})}/\hat{\sigma}_B$, which reveals that the
correction dedicated by the gluino exchange contribution is independent on
$m_{\tilde{g}}$, when the mass of gluino becomes heavy.

  The dependence of the corrections $\delta \hat{\sigma}^{(\tilde{g}
+ \tilde{t})}$ on the mixing angle $cos \theta_{\tilde{t}}$ for
$\sqrt{\hat{s}} =500~GeV$, $m_{\tilde{t}_1}=200~GeV$, $m_{\tilde{t}_2}=550~GeV$
and $m_{\tilde{g}}=250~GeV$
is shown in Fig.6. The combined correction
$\delta \hat{\sigma}^{(\tilde{g} + \tilde{t})}/\hat{\sigma}_B$ varies from
$-3 \%$ to $+5 \%$. From the figure one can see that the correction will reach
its peak values when $cos \theta_{\tilde{t}} = \pm 0.7$, namely when the SUSY
weak eigenstates $\tilde{t}_L$ and $\tilde{t}_R$ are mixed equivalently.

\noindent
\begin{flushleft} {\bf VI. Conclusion} \end{flushleft}

  In this work we have calculated the complete $O(\alpha_s)$ order SUSY QCD
corrections, including the conventional gluon contribution and the pure SUSY
particle contributions as well, to the $\tilde{t}_1 \bar{\tilde{t}}_1$ pair
production process via photon-photon fusion in the initial $e^{+} e^{-}$
collider.

  From the results of numerical evaluation, it is expected
that the corrections
due to the pure SUSY particle $\tilde{t}_i ~(i=1,2)$ and $\tilde{g}$
contributions are smaller than those from the virtual gluon loop
diagrams. Among the pure SUSY particle contributions, those
due to the squark quartic interactions are much smaller than those due to the
gluino exchange with reasonable parameters. The gluonic correction
may reach $+50\%$ in the photon
fusion subprocess, when $\sqrt{\hat{s}}$ is around the
$\tilde{t}_1 \bar{\tilde{t}}_1$ pair production threshold, whereas the
corrections there contributed by $\tilde{t}_i$ and $\tilde{g}$ diagrams are
only about $+10\%$. The total QCD corrections to the subprocess will be most
significant when $\sqrt{\hat{s}}$ is around the pair production threshold.
For the cross section of the process
$e^{+} e^{-} \rightarrow \gamma \gamma \rightarrow \tilde{t}_1 \bar{\tilde{t}}_1$,
the SUSY QCD correction at one-loop level can reach $35\%$.
Furthermore, the corrections also show the decoupling with
the heavy $\tilde{g}$ and $\tilde{t}_2$ masses, and the dependence of the QCD
corrections on the mixing angle $\theta_{\tilde{t}}$ obviously.

One of the authors, Z.H. Yu, would like to thank the Institute
of Theoretical Physics of University Vienna for the warm hospitality
extended to him during his stay under the agreement the exchange
program(project number: IV.B.12).

\vskip 10mm
\begin{center} {\Large Appendix}\end{center}

    The form factors $f_{i}^{(g)}$ due to the gluon contribution can be
expressed as
$$
\begin{array} {lll}
f_{vir, 1}^{(g)} &=& \frac{g_s^2 C_F}{8 \pi^2} \left \{
       B_0[p_1 + p_2, m_{\tilde{t}_1}, m_{\tilde{t}_1}] - \right. \\
& &~~~~~ ( (m_g^2 C_0 + m_{\tilde{t}_1}^2 2 C_{11}) -
           2 (p_1 \cdot p_2) (2 C_0 + C_{11}) ) \\
& &~~~~~ \left. [p_1, -p_1 - p_2, m_g, m_{\tilde{t}_1}, m_{\tilde{t}_1}] \right \} \\
&-& \frac{g_s^2 C_F}{4 \pi^2} \left \{
       (B_0[p_1 - p_3, m_g, m_{\tilde{t}_1}] -
        2 C_{24}[p_1, -p_3, m_g, m_{\tilde{t}_1}, m_{\tilde{t}_1}] +
        C_{24}[-p_3, -p_4, m_{\tilde{t}_1}, m_{\tilde{t}_1}, m_{\tilde{t}_1}] + \right. \\
& &~~~~~  ( 2 (p_1 \cdot p_2) (2 D_{27} + D_{311}) +
            2 (p_1 - p_2) \cdot p_3 (D_{312} - D_{313}) -
            m_g^2 D_{27} - 2 m_{\tilde{t}_1}^2 D_{311} ) \\
& &~~~~~  [p_1, -p_3, -p_4, m_g, m_{\tilde{t}_1}, m_{\tilde{t}_1}, m_{\tilde{t}_1}]
          ) + \left. (p_3 \leftrightarrow p_4) \right \} \\
&+& 2 \delta Z^{(g)}
\end{array}
\eqno{(A1)}
$$

$$
\begin{array} {lll}
f_{vir, 2}^{(g)} &=& -\frac{g_s^2 C_F}{4 \pi^2} \left \{
        (2 C_0 + 3 C_{11} - C_{12} + C_{21} - C_{23})
        [p_1, -p_3, m_g, m_{\tilde{t}_1}, m_{\tilde{t}_1}] \right. \\

& & - (C_{12} + C_{23})
      [-p_3, -p_4, m_{\tilde{t}_1}, m_{\tilde{t}_1}, m_{\tilde{t}_1}] \\

& & + ( 2 (D_{312} + D_{313} - D_{27} - 2 D_{311}) + \\
& &~~~~~m_g^2 (D_{11} - D_{12} + D_{21} - D_{24} - D_{25} + D_{26}) + \\
& &~~~~~2 m_{\tilde{t}_1}^2 (D_{21} - D_{24} + D_{31} + D_{310} -
                               D_{34} - D_{35}) + \\
& &~~~~~2 (p_1 \cdot p_2) (D_{34} + D_{35} - 2 D_{11} + 2 D_{12} -
                             3 D_{21} + 3 D_{24} + \\
& &~~~~~~~~~~~~~             2 D_{25} - 2 D_{26} - D_{31} - D_{310}) + \\
& &~~~~~2 (p_1 - p_2) \cdot p_3 (D_{22} - D_{24} + D_{25} - D_{26} -
                                   D_{34} + D_{35} + \\
& &~~~~~~~~~~~~~                   D_{36} - D_{37} - D_{38} + D_{39})) \\
& &~~~~~ \left.
[p_1, -p_3, -p_4, m_g, m_{\tilde{t}_1}, m_{\tilde{t}_1}, m_{\tilde{t}_1}] \right \} \\
&+& (p_3 \leftrightarrow p_4)
\end{array}
\eqno{(A2)}
$$

$$
\begin{array} {lll}
f_{vir, 3}^{(g)} &=& \frac{g_s^2 C_F}{4 \pi^2} \frac{1}{(t - m_{\tilde{t}_1}^2)^2} \left \{
        A_0[m_{\tilde{t}_1}] + (m_g^2 B_0 + 4t \cdot (B_0 + B_1)) [p_1 - p_3, m_g, m_{\tilde{t}_1}] \right \} \\
&-& \frac{1}{(t - m_{\tilde{t}_1}^2)^2} \left \{
        t \cdot \delta Z^{(g)} - m_{\tilde{t}_1}^2 \delta Z^{(g)} - \delta M^{2(g)} \right \} \\
&-& \frac{g_s^2 C_F}{2 \pi^2} \frac{1}{t - m_{\tilde{t}_1}^2} \left \{
        (2 B_0 + B_1)[p_1, m_g, m_{\tilde{t}_1}] + (2 B_0 + B_1)[p_1 - p_3, m_g, m_{\tilde{t}_1}] + \right. \\
& &~~~~~~~~~~~~~~(m_g^2 (C_0 +  C_{11}) + 4 m_{\tilde{t}_1}^2 (C_0 + 2 C_{11} + C_{21}) - 4 C_{24} - \\
& &~~~~~~~~~~~~~~~~2 (p_1 \cdot p_3) (2 C_0 + 3 C_{11} + 2 C_{12} + C_{21} + 2 C_{23})) \\
& &~~~~~~~~~~~~~~  \left. [p_1, -p_3, m_g, m_{\tilde{t}_1}, m_{\tilde{t}_1}] \right \} \\
&+& \frac{2}{t - m_{\tilde{t}_1}^2} \delta Z^{(g)} \\
&+& \frac{g_s^2 C_F}{4 \pi^2} \left \{
        2 (C_{12} + C_{23})[p_1, -p_3, m_g, m_{\tilde{t}_1}, m_{\tilde{t}_1}] + \right. \\
& &~~~~~~~~2 (C_{12} + C_{23})[-p_3, -p_4, m_{\tilde{t}_1}, m_{\tilde{t}_1}, m_{\tilde{t}_1}] - \\
& &~~~~~~~~( 2 (2 D_{27} + D_{311} + D_{312} - D_{313}) + \\
& &~~~~~~~~~~~  m_g^2 (D_{26} - D_0 - D_{11} - D_{12} + D_{13} - D_{24}) + \\
& &~~~~~~~~~~~  2 m_{\tilde{t}_1}^2 (D_{25} - D_{11} - D_{21} - D_{24} + D_{310} - D_{34}) + \\
& &~~~~~~~~~~~  2 (p_1 \cdot p_2) (2 D_0 + 3 D_{11} + 2 D_{12} -
                                 2 D_{13} + D_{21} + 3 D_{24} - \\
& &~~~~~~~~~~~~~~~~                D_{25} - 2 D_{26} - D_{310} + D_{34}) + \\
& &~~~~~~~~~~~  2 (p_1 - p_2) \cdot p_3 (D_{12} - D_{13} + D_{22} + D_{23} +
                                      D_{24} - D_{25} - \\
& &~~~~~~~~~~~~~~~~                     2 D_{26} - D_{310} + D_{36} - D_{38} + D_{39}) )\\
& &~~~~~~~~~~~~  [p_1, -p_3, -p_4, m_g, m_{\tilde{t}_1}, m_{\tilde{t}_1}, m_{\tilde{t}_1}] - \\
& &~~~~~~~~  ( 2 (D_{311} - D_{312} + D_{313}) + m_g^2 (D_{26} - D_{25}) + 2 m_{\tilde{t}_1}^2 (D_{310} - D_{35}) + \\
& &~~~~~~~~~~    2 (p_1 \cdot p_2) (2 D_{25} - 2 D_{26} - D_{310} + D_{35}) + \\
& &~~~~~~~~~~    2 (p_1 - p_2) \cdot p_4 (D_{310} - D_{37} - D_{38} + D_{39}) ) \\
& &~~~~~~~~~~ \left. [p_1, -p_4, -p_3, m_g, m_{\tilde{t}_1}, m_{\tilde{t}_1}, m_{\tilde{t}_1}] \right \} \\
\end{array}
\eqno{(A3)}
$$

$$
f_{vir, 4}^{(g)} = f_3^{(g)}(\hat{t} \rightarrow \hat{u},
                      p_3 \leftrightarrow p_4)
\eqno{(A4)}
$$

$$
\begin{array} {lll}
f_{vir, 5}^{(g)} &=& -\frac{g_s^2 C_F}{4 \pi^2} \left \{
        (2 C_0 + 3 C_{11} - C_{12} + C_{21} - C_{23})
        [p_1, -p_3, m_g, m_{\tilde{t}_1}, m_{\tilde{t}_1}] \right. \\

& & - (C_{12} - C_{23})[-p_3, -p_4, m_{\tilde{t}_1}, m_{\tilde{t}_1}, m_{\tilde{t}_1}] \\

& & + ( m_g^2 (D_{13} + D_{26}) + 2 m_{\tilde{t}_1}^2 (D_{25} + D_{310}) -
        2 (D_{27} + D_{312} + D_{313}) + \\
& &~~~~~2 (p_1 - p_2) \cdot p_3 (D_{23} - D_{26} - D_{38} + D_{39}) - \\
& &~~~~~2 (p_1 \cdot p_2) (2 D_{13} + D_{25} + 2 D_{26} + D_{310})) \\
& &~~~\left. [p_1, -p_3, -p_4, m_g, m_{\tilde{t}_1}, m_{\tilde{t}_1}, m_{\tilde{t}_1}] \right \} \\
&+& (p_3 \leftrightarrow p_4)
\end{array}
\eqno{(A5)}
$$

    The form factors $f_{i}^{(\tilde{g})}$ due to the gluino exchange can be
expressed as
$$
\begin{array} {lll}
f_1^{(\tilde{g})} &=& \frac{g_s^2 C_F}{16 \pi^2} \left \{ d\cdot B_0[-p_4, m_t, m_t]  \right. \\
&~&  -~2 (d\cdot C_{24} + 2 m_{\tilde{g}}^2 C_0 + 2 m_{\tilde{t}_1}^2 C_{12} +
         2 (p_1 \cdot p_2) (C_0 + C_{12}) + \\
& &~~~~~ 2 (p_2 \cdot p_3) (C_0 + C_{11}) +
         2 (p_1 \cdot p_3) C_{12} + m_t m_{\tilde{g}}~sin 2 \theta_{\tilde{t}}~C_0 ) \\
& &~~~~~~ [-p_3, -p_4, m_t, m_t, m_t] \\
&~&  -~4 (m_{\tilde{g}}^2 (m_t^2 - m_{\tilde{g}}^2) D_0 - 2 (m_t^2 + m_{\tilde{g}}^2) D_{27} + \\
& &~~~~~  m_{\tilde{t}_1}^2 (m_t^2 D_{13} + m_{\tilde{g}}^2 (D_{13} - D_0 - 2 D_{11}) +
                   m_{\tilde{t}_1}^2 (2 D_{23} - D_{13} - 2 D_{25}) ) + \\
& &~~~~~  (p_1 \cdot p_2) (m_t^2 (D_{13} - D_{11}) - 4 D_{27} +
                    m_{\tilde{g}}^2 (D_{11} + D_{13}) + \\
& &~~~~~~~~~~~~~      m_{\tilde{t}_1}^2 (D_{11} - D_{13} + 2 D_{21} +
                            4 D_{23} - 4 D_{25}) ) + \\
& &~~~~~  (p_1 \cdot p_3) (m_t^2 (D_{11} - D_{13}) +
                    m_{\tilde{g}}^2 (D_{0} + D_{11} + 2 D_{12} - 3 D_{13}) + \\
& &~~~~~~~~~~~~~    m_{\tilde{t}_1}^2 (D_{13} - 2 D_{23} + 2 D_{26}) ) + \\
& &~~~~~  (p_2 \cdot p_3) (m_t^2 (D_{12} - 2 D_{13}) - m_{\tilde{g}}^2 (D_{0} + D_{12}) + \\
& &~~~~~~~~~~~~~    m_{\tilde{t}_1}^2 (2 D_{13} - D_{11} - D_{12} - 2 D_{23} -
                            2 D_{24} + 2 D_{25} + 2 D_{26}) ) + \\
& &~~~~~  2 (p_1 \cdot p_2)^2 (D_{23} - D_{25}) +
          2 (p_1 \cdot p_2) (p_2 \cdot p_3) (D_{26} - D_{23}) + \\
& &~~~~~  2 (p_1 \cdot p_2) (p_1 \cdot p_3) (D_{25} + D_{26} - D_{23} - D_{24}) + \\
& &~~~~~  2 (p_1 \cdot p_3) (p_2 \cdot p_3) (D_{12} - D_{13} + D_{22} + D_{23} - 2 D_{26}) + \\
& &~~~~~  m_t m_{\tilde{g}}~sin 2 \theta_{\tilde{t}}~((m_t^2 - m_{\tilde{g}}^2) D_{0} - 4 D_{27} +
                             m_{\tilde{t}_1}^2 (2 D_{13} - D_{0} - 2 D_{11}) + \\
& &~~~~~~~~~~~~~    2 (p_1 \cdot p_2) D_{13} - (p_2 \cdot p_3) (D_{0} + 2 D_{13}) + \\
& &~~~~~~~~~~~~~    (p_1 \cdot p_3) (D_{0} + 2 D_{11} + 2 D_{12} - 4 D_{13}) ) ) \\
& &~~~~~  \left. [p_1, -p_3, -p_4, m_{\tilde{g}}, m_t, m_t, m_t] \right\} \\
&+& (p_3 \leftrightarrow p_4) + 2~\delta Z^{(\tilde{g})}
\end{array}
\eqno{(A6)}
$$

$$
\begin{array} {lll}
f_2^{(\tilde{g})} &=& \frac{g_s^2 C_F}{4 \pi^2} \left \{
      (C_{12} + 2 C_{23})[-p_3, -p_4, m_t, m_t, m_t] \right. \\
&~& -~(m_t^2 (D_{11} - 2 D_{12} + D_{13} + 2 D_{21} - 2 D_{24} - 2 D_{25} + 2 D_{26}) + \\
&~&~~~~~ m_{\tilde{g}}^2 (D_{0} + 3 D_{11} - 2 D_{12} - D_{13} +
         2 D_{21} - 2 D_{24} - 2 D_{25} + 2 D_{26}) + \\
&~&~~~~~ m_{\tilde{t}_1}^2 (D_{13} - 2 D_{23} + 2 D_{25}) + \\
&~&~~~~~ 2 ((p_1 \cdot p_2) - (p_2 \cdot p_3))
         (D_{13} - D_{12} - D_{23} - D_{24} + D_{25} + D_{26}) + \\
&~&~~~~~ m_t m_{\tilde{g}}~sin 2 \theta_{\tilde{t}}
         (D_{0} + 4 (D_{11} - D_{12} + D_{21} - D_{24} - D_{25} + D_{26})) ) \\
&~&~~ \left.  [p_1, -p_3, -p_4, m_{\tilde{g}}, m_t, m_t, m_t] \right\} \\
&+& (p_3 \leftrightarrow p_4)
\end{array}
\eqno{(A7)}
$$

$$
\begin{array} {lll}
f_3^{(\tilde{g})} &=& \frac{d \cdot g_s^2 C_F}{4 \pi^2 (\hat{t} - m_{\tilde{t}_1}^2)^2}
         \left \{ ((m_{\tilde{g}}^2 + m_t m_{\tilde{g}}~sin 2 \theta_{\tilde{t}}) B_0 +
         \hat{t} \cdot B_1)[p_1 - p_3, m_{\tilde{g}}, m_t] - A_0[m_t] \right \} \\
&+& \frac{g_s^2 C_F}{4 \pi^2 (\hat{t} - m_{\tilde{t}_1}^2)}
     \left \{ d \cdot B_{0}[-p_3, m_t, m_t] +
              4 \cdot (2 (p_1, p_3) C_{12} - m_t^2 C_{11} - \right. \\
& &~~~~~  m_{\tilde{g}}^2 \cdot (2 C_{0} + C_{11}) - m_{\tilde{t}_1}^2 C_{11} - \\
& &~~~~~ \left. 2 m_t m_{\tilde{g}}~sin 2 \theta_{\tilde{t}}~(C_{0} + C_{11}) )
        [p_1, -p_3, m_{\tilde{g}}, m_t, m_t] \right \} \\
&-& \frac{4}{(\hat{t} - m_{\tilde{t}_1}^2)^2}
    (\hat{t} \cdot \delta Z^{(\tilde{g})} - \delta M^{2(\tilde{g})} -
     m_{\tilde{t}_1}^2 \cdot \delta Z^{(\tilde{g})})
 + \frac{8}{\hat{t} - m_{\tilde{t}_1}^2} \delta Z^{(\tilde{g})} \\
&+& \frac{g_s^2 C_F}{4 \pi^2} \left \{
  (C_{0} + C_{11} + 3 C_{12} + 4 C_{23})[-p_3, -p_4, m_t, m_t, m_t] \right. \\
& &~~~ -(2 m_t^2 (D_{13} - D_{11} - D_{12} - D_{24} + D_{26}) - m_{\tilde{t}_1}^2 D_{11} + \\
& &~~~~~~~  m_{\tilde{g}}^2 (2 (D_{13} - D_{11} - D_{12} - D_{24} + D_{26}) - 3 D_{0}) + \\
& &~~~~~~~  2 (p_1 \cdot p_2) (D_{13} - D_{12} - D_{24} + D_{26}) + 2 (p_1\cdot p_3) (D_{12} - D_{13}) +\\
& &~~~~~~~  m_t m_{\tilde{g}}~sin 2 \theta_{\tilde{t}}~(4 (D_{13} - D_{11} - D_{12} - D_{24} + D_{26}) - 3 D_{0})) \\
& &~~~~~~ [p_1, -p_3, -p_4, m_{\tilde{g}}, m_t, m_t, m_t] \\
& &~~~ -(m_{\tilde{g}}^2 (D_{0} - D_{11} + D_{12} - D_{13} - 2 D_{25} + 2 D_{26}) + 4 D_{27} + \\
& &~~~~~~~  m_t^2 (D_{11} - D_{12} + D_{13} - 2 D_{25} + 2 D_{26}) +\\
& &~~~~~~~  m_{\tilde{t}_1}^2 (D_{12} - D_{13} - 2 D_{21} - 2 D_{23} + 2 D_{24} + 2 D_{25} - 2 D_{26}) +\\
& &~~~~~~~  2 (p_1 \cdot p_2) (D_{25} - D_{23}) + 2 (p_2 \cdot p_4) (D_{23} - D_{26}) + \\
& &~~~~~~~  2 (p_1 \cdot p_4) (D_{13} - D_{12} - D_{22} + D_{24} - D_{25} + D_{26}) +\\
& &~~~~~~~  m_t m_{\tilde{g}}~sin 2 \theta_{\tilde{t}}~(D_{0} - 4 D_{25} + 4 D_{26}))\\
& &~~~~~~ \left. [p_1, -p_4, -p_3, m_{\tilde{g}}, m_t, m_t, m_t] \right \}
\end{array}
\eqno{(A8)}
$$

$$
f_4^{(\tilde{g})} = f_3^{(\tilde{g})}(\hat{t} \rightarrow \hat{u},
                     p_3 \leftrightarrow p_4)
\eqno{(A9)}
$$

$$
\begin{array} {lll}
f_5^{(\tilde{g})} &=& \frac{g_s^2 C_F}{4 \pi^2} \left \{
   (C_0 + C_{11} + 2 C_{12} + 2 C_{23})[-p_3, -p_4, m_t, m_t, m_t] \right. \\
&~&  -~(m_{\tilde{t}_1}^2 (D_{11} + D_{12} - 2 D_{13} + 2 D_{24} - 2 D_{26}) + \\
& &~~~~~     m_{\tilde{g}}^2 (D_{0} + D_{12} + 2 D_{13} + 2 D_{26}) + \\
& &~~~~~     m_t^2 (2 D_{13} - D_{12} + 2 D_{26}) + \\
& &~~~~~     2 (p_1\cdot p_3) (D_{13} - D_{12} - D_{22} + D_{26}) + \\
& &~~~~~     m_t m_{\tilde{g}}~sin 2 \theta_{\tilde{t}}~(D_{0} + 4 D_{13} + 4 D_{26})) \\
& &~~~~ \left.   [p_1, -p_3, -p_4, m_{\tilde{g}}, m_t, m_t, m_t] \right \} \\
&+& (p_3 \leftrightarrow p_4)
\end{array}
\eqno{(A10)}
$$

    The factors arising from the contribution of squark quartic interactions, are
given as follow
$$
\begin{array} {lll}
f_1^{(\tilde{t})} &=&
    \frac{g_s^2}{6 \pi^2} cos^2 2 \theta_{\tilde{t}} \left\{ 4 \cdot C_{24}
      [-p_3, p_3 + p_4, m_{\tilde{t}_1}, m_{\tilde{t}_1}, m_{\tilde{t}_1}]
   - B_0 [p_1 + p_2, m_{\tilde{t}_1}, m_{\tilde{t}_1}] \right\} \\
&+& (cos \rightarrow sin, m_{\tilde{t}_1} \rightarrow m_{\tilde{t}_2})
\end{array}
\eqno{(A11)}
$$

$$
\begin{array} {lll}
f_2^{(\tilde{t})} &=&
    \frac{2 g_s^2}{3 \pi^2} cos^2 2 \theta_{\tilde{t}}~(C_{23} - C_{22})
         [-p_3, p_3 + p_4, m_{\tilde{t}_1}, m_{\tilde{t}_1}, m_{\tilde{t}_1}] \\
&+& (cos \rightarrow sin, m_{\tilde{t}_1} \rightarrow m_{\tilde{t}_2})
\end{array}
\eqno{(A12)}
$$

$$
\begin{array} {lll}
f_3^{(\tilde{t})} &=&
    \frac{2 g_s^2}{3 \pi^2} cos^2 2 \theta_{\tilde{t}}~(C_{23} - C_{22})
     [-p_3, p_3 + p_4, m_{\tilde{t}_1}, m_{\tilde{t}_1}, m_{\tilde{t}_1}] \\
&+& (cos \rightarrow sin, m_{\tilde{t}_1} \rightarrow m_{\tilde{t}_2})
\end{array}
\eqno{(A13)}
$$

$$
f_4^{(\tilde{t})} = f_3^{(\tilde{t})}(\hat{t} \rightarrow \hat{u})
\eqno{(A14)}
$$

$$
f_5^{(\tilde{t})} = f_2^{(\tilde{t})}
\eqno{(A15)}
$$

  The factors shown in Eq.(3.11b) have the expressions as

\begin{eqnarray*}
f_{real}^{ \hat{t} \hat{t}} &=& 2 \cdot I_2 - m_{\tilde{t}_1}^2 \cdot (I_{11} + I_{22}) +
         2 (p_1 \cdot p_2) \cdot I_{12} \\
&-&      \frac{2 m_{\tilde{t}_1}^2}{(p_1 \cdot p_2)} (I_1 + I_2) +
         \frac{2 m_{\tilde{t}_1}^2}{(p_1 \cdot p_3)} I_1 +
         \frac{2 (p_3 - p_1) \cdot p_2}{(p_1 \cdot p_3)} I_2 \\
&-&      \frac{2 m_{\tilde{t}_1}^2}{(p_1 \cdot p_2)^2} -
                 \frac{m_{\tilde{t}_1}^2}{(p_1 \cdot p_3)^2} +
                 \frac{2 m_{\tilde{t}_1}^2}{(p_1 \cdot p_2) (p_1 \cdot p_3)} +
                 \frac{2 (p_2 \cdot p_3)}{(p_1 \cdot p_2) (p_1 \cdot p_3)}
\end{eqnarray*}
$$
\eqno{(A16)}
$$

$$
f_{real}^{ \hat{u} \hat{u}} = f_{real}^{ \hat{t} \hat{t}} (p_3 \rightarrow p_4)
\eqno{(A17)}
$$

$$
f_{real}^{ \hat{q} \hat{q}} = 2 (p_1 \cdot p_2) I_{12} - m_{\tilde{t}_1}^2 (I_{11} + I_{22})
\eqno{(A18)}
$$

\begin{eqnarray*}
f_{real}^{ \hat{t} \hat{u}} &=& 2 \cdot I_2
  - m_{\tilde{t}_1}^2 \cdot (I_{11} + I_{22})
  + 2 (p_1 \cdot p_2) \cdot I_{12} \\
&-& \frac{2 m_{\tilde{t}_1}^2}{(p_1 \cdot p_2)} (I_1 + I_2)
  + \frac{m_{\tilde{t}_1}^2}{(p_1 \cdot p_3)} I_1
  + \frac{m_{\tilde{t}_1}^2}{(p_1 \cdot p_4)} I_1
  + \frac{(p_3 - p_1) \cdot p_2}{(p_1 \cdot p_3)} I_2
  + \frac{(p_4 - p_1) \cdot p_2}{(p_1 \cdot p_4)} I_2 \\
&-& \frac{2 m_{\tilde{t}_1}^2}{(p_1 \cdot p_2)^2}
  + \frac{m_{\tilde{t}_1}^2}{(p_1 \cdot p_2) (p_1 \cdot p_3)}
  + \frac{m_{\tilde{t}_1}^2}{(p_1 \cdot p_2) (p_1 \cdot p_4)} \\
&-& \frac{m_{\tilde{t}_1}^2}{(p_1 \cdot p_3) (p_1 \cdot p_4)}
  + \frac{(p_2 \cdot p_3)}{(p_1 \cdot p_2) (p_1 \cdot p_3)}
  + \frac{(p_2 \cdot p_4)}{(p_1 \cdot p_2) (p_1 \cdot p_4)}
  - \frac{(p_3 \cdot p_4)}{(p_1 \cdot p_3) (p_1 \cdot p_4)}
\end{eqnarray*}
$$
\eqno{(A19)}
$$

\begin{eqnarray*}
f_{real}^{ \hat{q} \hat{t}} &=& I_2
  - m_{\tilde{t}_1}^2 \cdot (I_{11} + I_{22})
  + 2 (p_1 \cdot p_2) \cdot I_{12} \\
&+& \frac{m_{\tilde{t}_1}^2}{(p_1 \cdot p_3)} I_1
  - \frac{m_{\tilde{t}_1}^2}{(p_1 \cdot p_2)} I_1
  - \frac{m_{\tilde{t}_1}^2}{(p_1 \cdot p_2)} I_2
  + \frac{(p_3 - p_1) \cdot p_2}{(p_1 \cdot p_3)} I_2
\end{eqnarray*}
$$
\eqno{(A20)}
$$

$$
f_{real}^{ \hat{q} \hat{u}} = f_{real}^{ \hat{q} \hat{t}}(p_3 \rightarrow p_4)
\eqno{(A21)}
$$

with

$$
I_{ab...} = \frac{1}{(p_a \cdot k)(p_b \cdot k)...} \hskip 10mm (a,b=1,2)
\eqno{(A22)}
$$

\vskip 10mm
\begin{flushleft} {\bf Figure Captions} \end{flushleft}

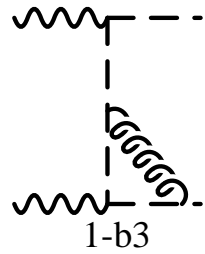
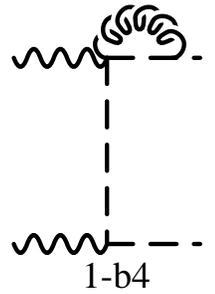
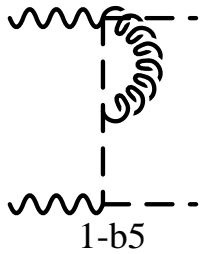
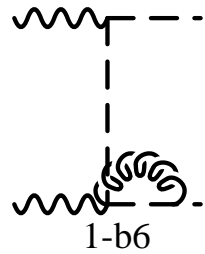
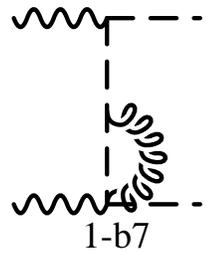
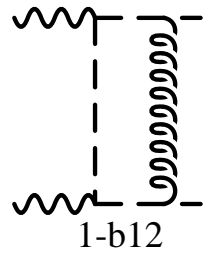
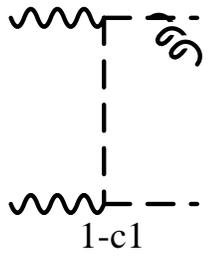
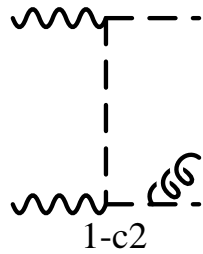
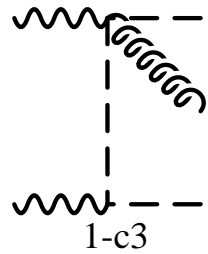
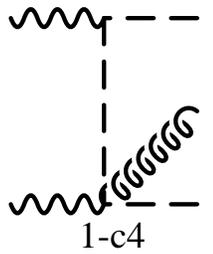
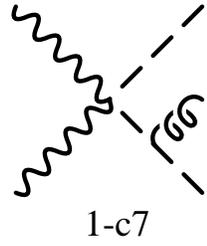
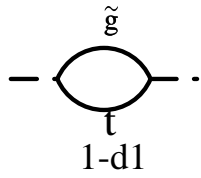
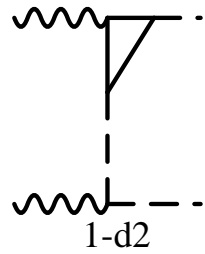
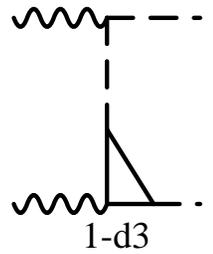
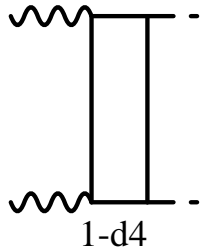
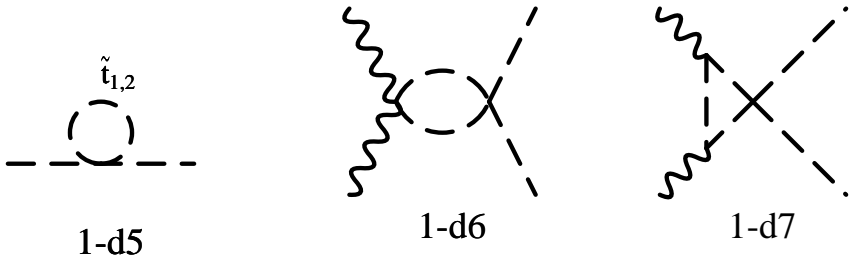
{\bf Fig.1} Feynman diagrams including the contributions from
        the tree-level and next-to-leading order in Supersymmetric
        QCD for $\gamma \gamma \rightarrow \tilde{t}_1 \bar{\tilde{t}}_1$
        process:
(1-a) tree level diagrams; (1-b) corrections due to gluon contribution;
(1-c) soft gluon emission; (1-d) corrections due to the pure supersymmetric
particle $\tilde{g}$ and $\tilde{t}_i$. The diagrams by exchanging initial
photons are not shown here.

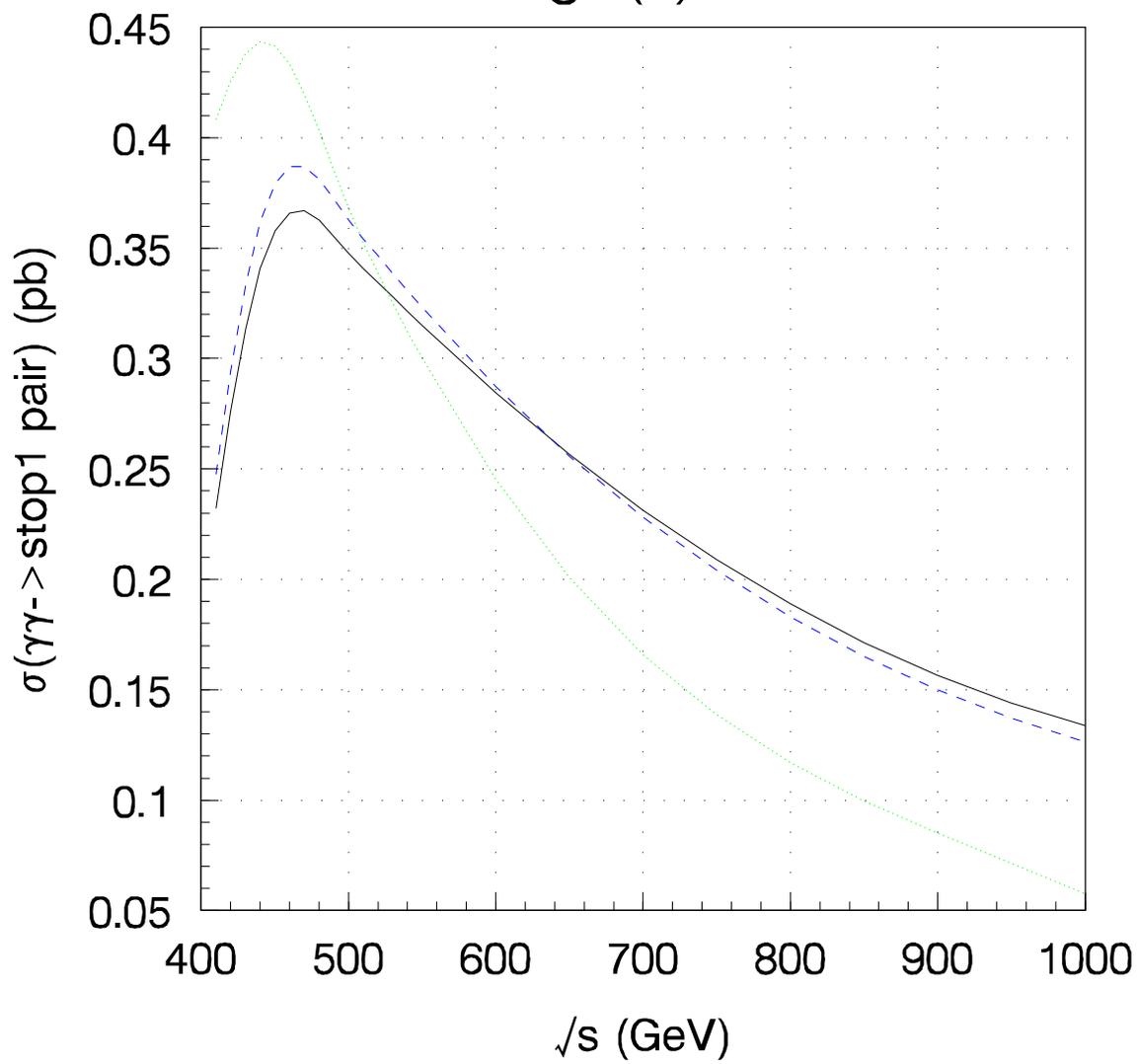
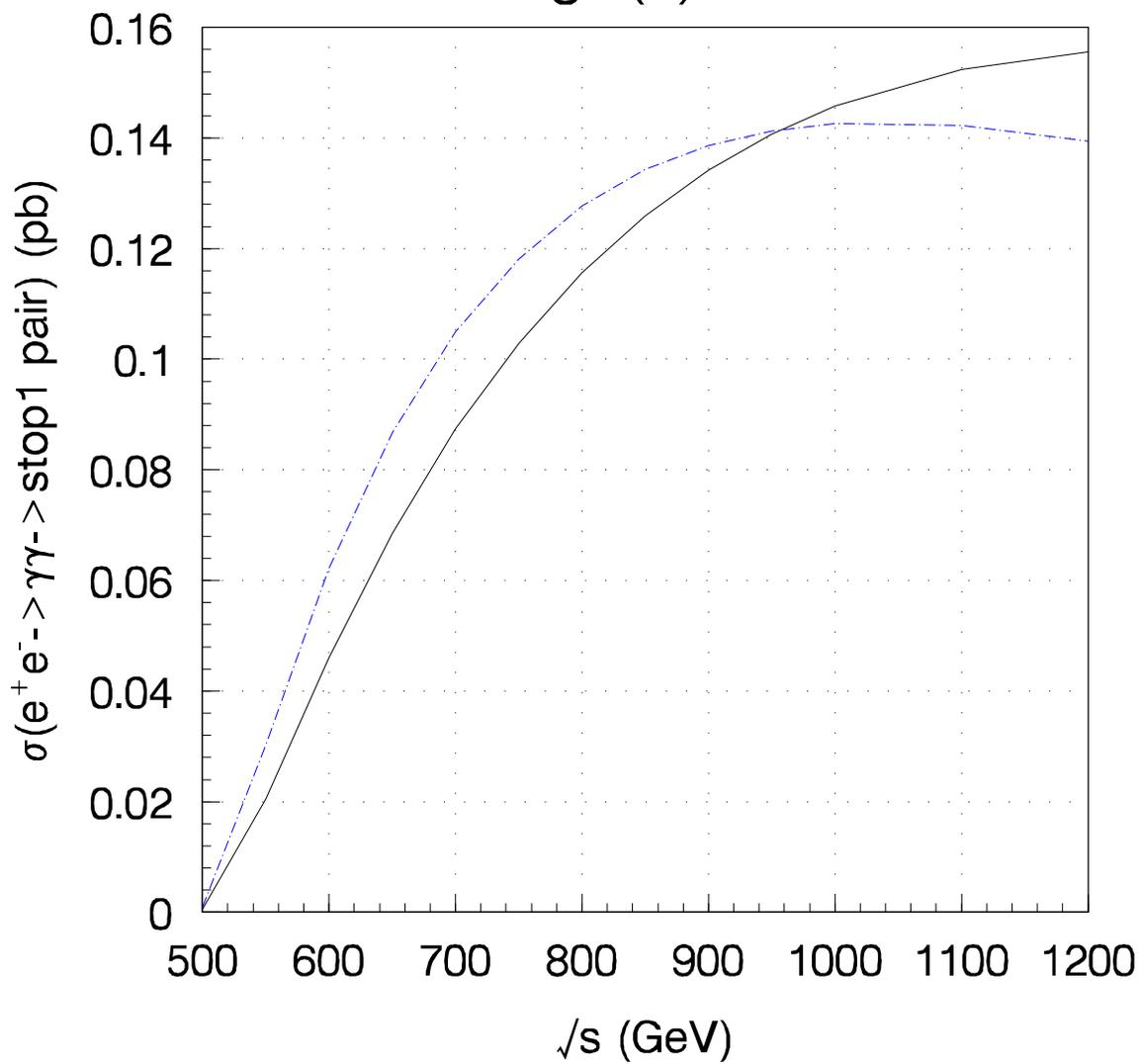
{\bf Fig.2} When $m_{\tilde{g}}=250~GeV$, $m_{\tilde{t}_1}=200~GeV$,
$m_{\tilde{t}_2}=550~GeV$ and $cos \theta_{\tilde{t}} = 0.7$:
(a) the cross section $\hat{\sigma}(\gamma \gamma \rightarrow \tilde{t}_1
    \bar{\tilde{t}}_1)$ as a function of $\sqrt{\hat{s}}$,
    solid-line for tree-level contribution $\hat{\sigma}_{B}$,
    dotted-line for the conventional QCD correction $\hat{\sigma}^{(g)}$,
    dashed-line for the pure supersymmetric particle QCD contribution
                    $\hat{\sigma}^{(\tilde{g}+\tilde{t})}$;
(b) the cross section $\sigma(e^{+}e^{-} \rightarrow \gamma\gamma \rightarrow
\tilde{t}_1\bar{\tilde{t}}_1)$ as a function of $\sqrt{s}$,
    solid-line for tree level contribution $\sigma_{B}$,
    dash-dotted-line for the overall QCD correction
                         $\sigma^{(g+\tilde{g}+\tilde{t})}$.

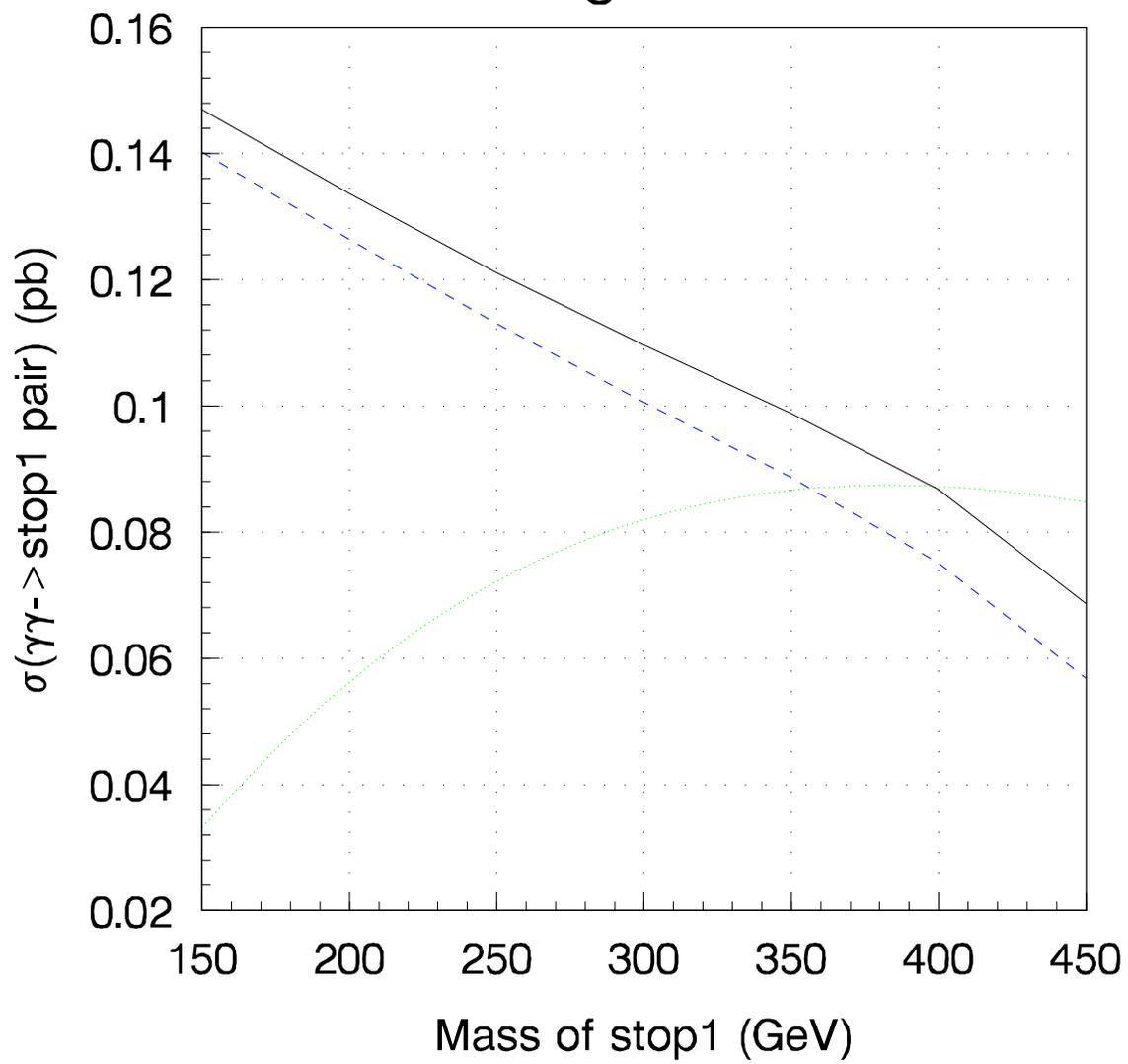
{\bf Fig.3} The cross section $\hat{\sigma}$ as a function of
$m_{\tilde{t}_1}$, when $\sqrt{\hat{s}}=1000~GeV$, $m_{\tilde{t}_2}=550~GeV$,
$m_{\tilde{g}}=250~GeV$ and $cos \theta_{\tilde{t}} = 0.7$:
    solid-line for tree-level contribution $\hat{\sigma}_{B}$,
    dotted-line for the conventional QCD correction $\hat{\sigma}^{(g)}$,
    dashed-line for the pure supersymmetric particle QCD contribution
                    $\hat{\sigma}^{(\tilde{g}+\tilde{t})}$.

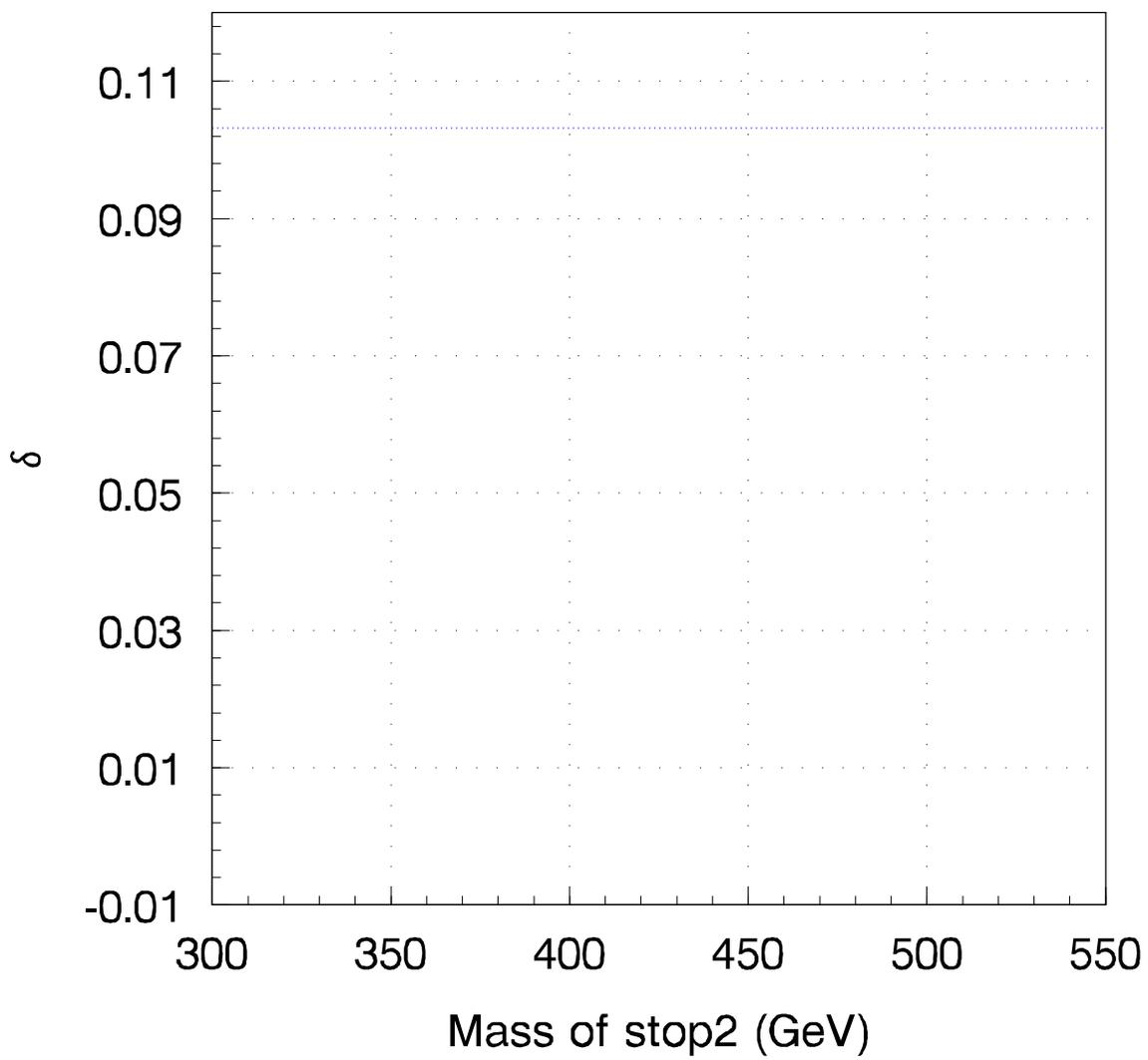
{\bf Fig.4} The relative correction $\delta \hat{\sigma}/\hat{\sigma}_B$ as a
function of $m_{\tilde{t}_2}$, when $\sqrt{\hat{s}}=500~GeV$, $m_{\tilde{t}_1}=200~GeV$,
$m_{\tilde{g}}=250~GeV$ and $cos \theta_{\tilde{t}} = 0.7$:
    dotted-line for the contribution
                         $\delta \hat{\sigma}^{(g + \tilde{g})}/\hat{\sigma}_B$,
    dashed-line for the QCD correction due to squark quartic interactions
                         $\delta \hat{\sigma}^{(\tilde{t})}/\hat{\sigma}_B$.

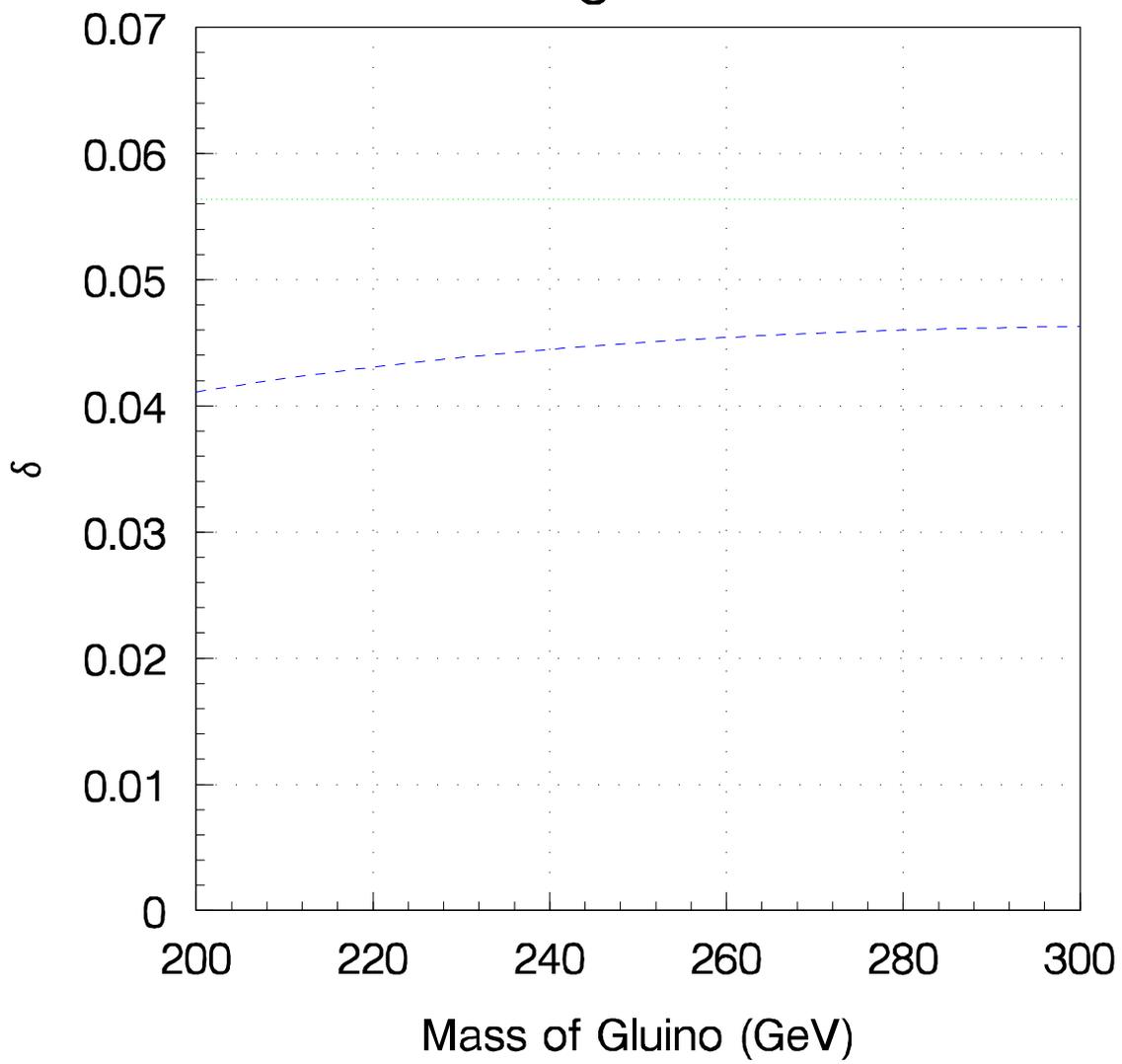
{\bf Fig.5} The relative correction $\delta \hat{\sigma}/\hat{\sigma}_B$ as a
function of
$m_{\tilde{g}}$, when $\sqrt{\hat{s}}=500~GeV$, $m_{\tilde{t}_1}=200~GeV$,
$m_{\tilde{t}_2}=550~GeV$ and $cos \theta_{\tilde{t}} = 0.7$:
    dotted-line for the contribution
                         $\delta \hat{\sigma}^{(g + \tilde{t})}/\hat{\sigma}_B$,
    dashed-line for the QCD correction due to the gluino exchange
                         $\delta \hat{\sigma}^{(\tilde{g})}/\hat{\sigma}_B$.

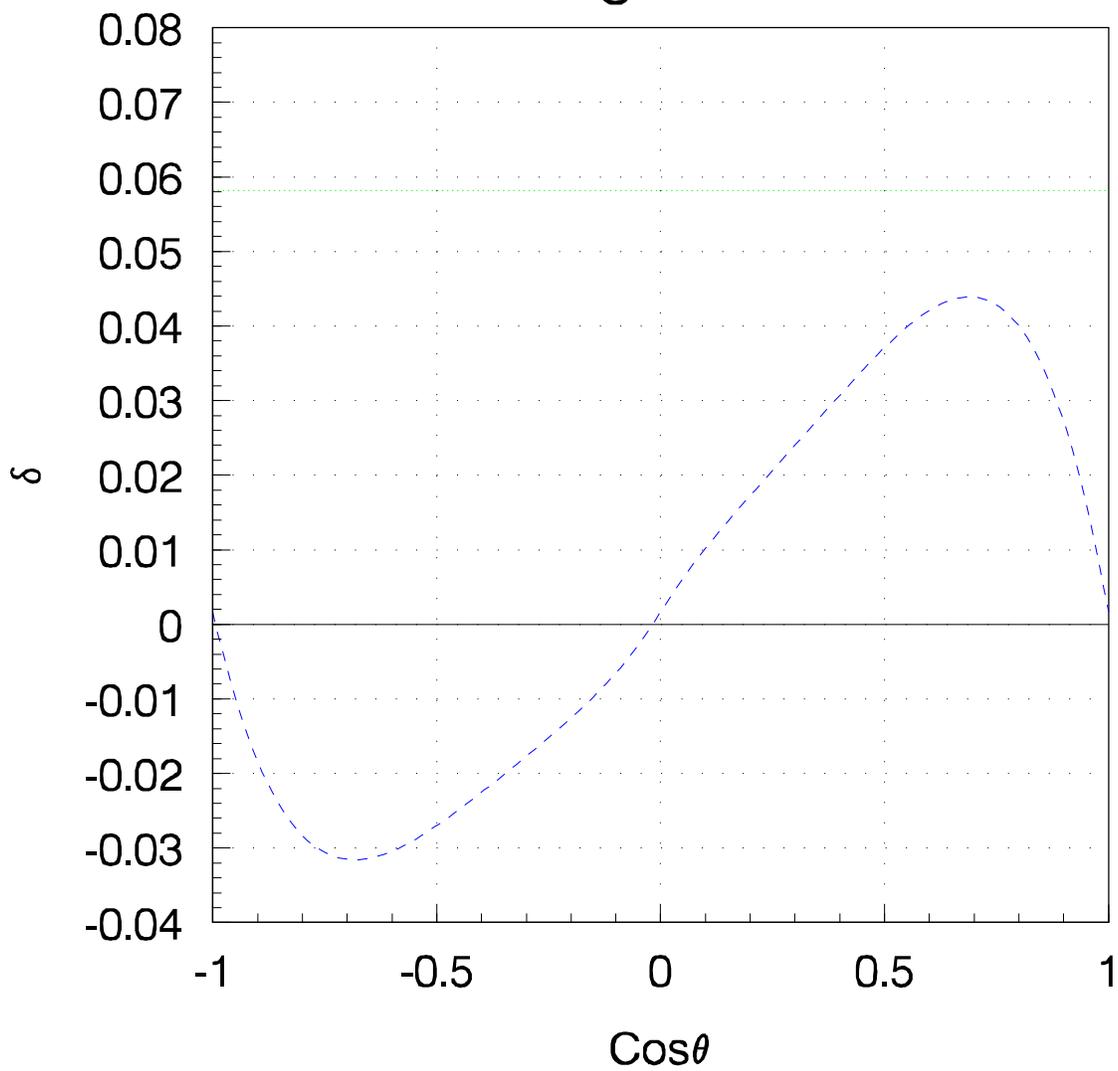
{\bf Fig.6} The relative correction $\delta \hat{\sigma}/\hat{\sigma}_B$ as a
function of
$cos \theta_{\tilde{t}}$, when $\sqrt{\hat{s}}=500~GeV$,
$m_{\tilde{t}_1}=200~GeV$, $m_{\tilde{t}_2}=550~GeV$ and
$m_{\tilde{g}}=250~GeV$:
    dotted-line for the gluon contribution
                         $\delta \hat{\sigma}^{(g)}/\hat{\sigma}_B$,
    dash-line for the pure supersymmetric particle contribution
                    $\delta \hat{\sigma}^{(\tilde{g}+\tilde{t})}/\hat{\sigma}_B$.

\end{large}
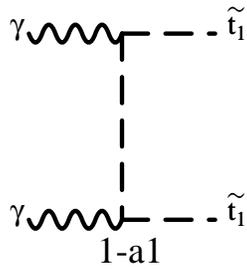
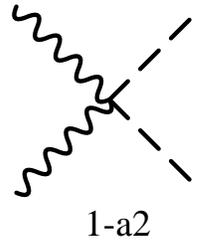
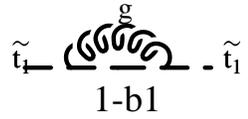
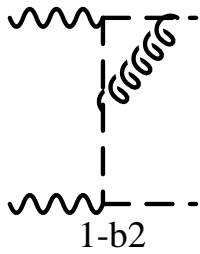
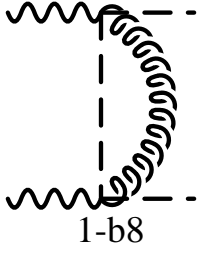
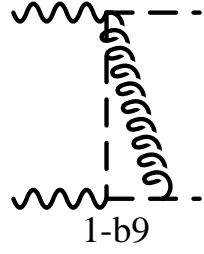
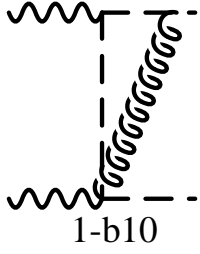
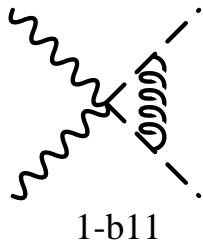
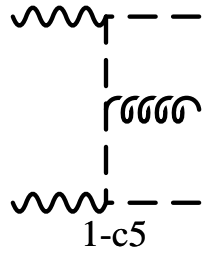
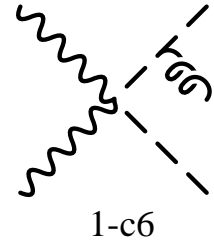
\end{document}